\PassOptionsToPackage{unicode}{hyperref}
\PassOptionsToPackage{hyphens}{url}
\PassOptionsToPackage{dvipsnames,svgnames,x11names}{xcolor}
\documentclass[
]{article}
\usepackage{amsmath,amssymb}
\usepackage{lmodern}
\usepackage{iftex}
\ifPDFTeX
  \usepackage[T1]{fontenc}
  \usepackage[utf8]{inputenc}
  \usepackage{textcomp} 
\else 
  \usepackage{unicode-math}
  \defaultfontfeatures{Scale=MatchLowercase}
  \defaultfontfeatures[\rmfamily]{Ligatures=TeX,Scale=1}
\fi
\IfFileExists{upquote.sty}{\usepackage{upquote}}{}
\IfFileExists{microtype.sty}{
  \usepackage[]{microtype}
  \UseMicrotypeSet[protrusion]{basicmath} 
}{}
\makeatletter
\@ifundefined{KOMAClassName}{
  \IfFileExists{parskip.sty}{%
    \usepackage{parskip}
  }{
    \setlength{\parindent}{0pt}
    \setlength{\parskip}{6pt plus 2pt minus 1pt}}
}{
  \KOMAoptions{parskip=half}}
\makeatother
\usepackage{xcolor}
\NewDocumentCommand\citeproctext{}{}
\NewDocumentCommand\citeproc{mm}{%
  \begingroup\def\citeproctext{#2}\cite{#1}\endgroup}
\makeatletter
 \let\@cite@ofmt\@firstofone
 \def\@biblabel#1{}
 \def\@cite#1#2{{#1\if@tempswa , #2\fi}}
\makeatother
\newlength{\cslhangindent}
\newlength{\csllabelwidth}
\newenvironment{CSLReferences}[2] 
 {\begin{list}{}{%
  \setlength{\itemindent}{0pt}
  \setlength{\leftmargin}{0pt}
  \setlength{\parsep}{0pt}
  \ifodd #1
   \setlength{\leftmargin}{\cslhangindent}
   \setlength{\itemindent}{-1\cslhangindent}
  \fi
  \setlength{\itemsep}{#2\baselineskip}}}
 {\end{list}}
\usepackage{calc}

\ifLuaTeX
\usepackage[bidi=basic]{babel}
\else
\usepackage[bidi=default]{babel}
\fi
\babelprovide[main,import]{american}

\def\languageshorthands#1{}
\ifLuaTeX
  \usepackage{selnolig}  
\fi
\IfFileExists{bookmark.sty}{\usepackage{bookmark}}{\usepackage{hyperref}}
\IfFileExists{xurl.sty}{\usepackage{xurl}}{} 
\hypersetup{
  pdftitle={pynxtools: A Python framework for generating and validating
NeXus files in experimental data workflows},
  pdfauthor={Sherjeel Shabih, Lukas Pielsticker, Florian Dobener, Andrea
Albino, Theodore Chang, Carola Emminger, Lev Ginzburg, Ron Hildebrandt,
Markus Kühbach, Rubel Mozumder, Tommaso Pincelli, Martin Aeschlimann,
Marius Grundmann, Walid Hetaba, Carlos-Andres Palma, Laurenz Rettig,
Markus Scheidgen, José Antonio Márquez Prieto, Claudia Draxl, Sandor
Brockhauser, Christoph Koch, Heiko B. Weber},
  pdflang={en-US},
  colorlinks=true,
  linkcolor={Maroon},
  filecolor={Maroon},
  citecolor={Blue},
  urlcolor={Blue},
  pdfcreator={LaTeX via pandoc}}

\title{pynxtools: A Python framework for generating and validating NeXus
files in experimental data workflows}

\definecolor{c53baa1}{RGB}{83,186,161}
\definecolor{c202826}{RGB}{32,40,38}
\def \rorglobalscale {0.1}
\newcommand{\rorlogo}{%
\begin{tikzpicture}[y=1cm, x=1cm, yscale=\rorglobalscale,xscale=\rorglobalscale, every node/.append style={scale=\rorglobalscale}, inner sep=0pt, outer sep=0pt]
  \begin{scope}[even odd rule,line join=round,miter limit=2.0,shift={(-0.025, 0.0216)}]
    \path[fill=c53baa1,nonzero rule,line join=round,miter limit=2.0] (1.8164, 3.012) -- (1.4954, 2.5204) -- (1.1742, 3.012) -- (1.8164, 3.012) -- cycle;
    \path[fill=c53baa1,nonzero rule,line join=round,miter limit=2.0] (3.1594, 3.012) -- (2.8385, 2.5204) -- (2.5172, 3.012) -- (3.1594, 3.012) -- cycle;
    \path[fill=c53baa1,nonzero rule,line join=round,miter limit=2.0] (1.1742, 0.0669) -- (1.4954, 0.5588) -- (1.8164, 0.0669) -- (1.1742, 0.0669) -- cycle;
    \path[fill=c53baa1,nonzero rule,line join=round,miter limit=2.0] (2.5172, 0.0669) -- (2.8385, 0.5588) -- (3.1594, 0.0669) -- (2.5172, 0.0669) -- cycle;
    \path[fill=c202826,nonzero rule,line join=round,miter limit=2.0] (3.8505, 1.4364).. controls (3.9643, 1.4576) and (4.0508, 1.5081) .. (4.1098, 1.5878).. controls (4.169, 1.6674) and (4.1984, 1.7642) .. (4.1984, 1.8777).. controls (4.1984, 1.9719) and (4.182, 2.0503) .. (4.1495, 2.1132).. controls (4.1169, 2.1762) and (4.0727, 2.2262) .. (4.0174, 2.2635).. controls (3.9621, 2.3006) and (3.8976, 2.3273) .. (3.824, 2.3432).. controls (3.7505, 2.359) and (3.6727, 2.367) .. (3.5909, 2.367) -- (2.9676, 2.367) -- (2.9676, 1.8688).. controls (2.9625, 1.8833) and (2.9572, 1.8976) .. (2.9514, 1.9119).. controls (2.9083, 2.0164) and (2.848, 2.1056) .. (2.7705, 2.1791).. controls (2.6929, 2.2527) and (2.6014, 2.3093) .. (2.495, 2.3487).. controls (2.3889, 2.3881) and (2.2728, 2.408) .. (2.1468, 2.408).. controls (2.0209, 2.408) and (1.905, 2.3881) .. (1.7986, 2.3487).. controls (1.6925, 2.3093) and (1.6007, 2.2527) .. (1.5232, 2.1791).. controls (1.4539, 2.1132) and (1.3983, 2.0346) .. (1.3565, 1.9436).. controls (1.3504, 2.009) and (1.3351, 2.0656) .. (1.3105, 2.1132).. controls (1.2779, 2.1762) and (1.2338, 2.2262) .. (1.1785, 2.2635).. controls (1.1232, 2.3006) and (1.0586, 2.3273) .. (0.985, 2.3432).. controls (0.9115, 2.359) and (0.8337, 2.367) .. (0.7519, 2.367) -- (0.1289, 2.367) -- (0.1289, 0.7562) -- (0.4837, 0.7562) -- (0.4837, 1.4002) -- (0.6588, 1.4002) -- (0.9956, 0.7562) -- (1.4211, 0.7562) -- (1.0118, 1.4364).. controls (1.1255, 1.4576) and (1.2121, 1.5081) .. (1.2711, 1.5878).. controls (1.2737, 1.5915) and (1.2761, 1.5954) .. (1.2787, 1.5991).. controls (1.2782, 1.5867) and (1.2779, 1.5743) .. (1.2779, 1.5616).. controls (1.2779, 1.4327) and (1.2996, 1.3158) .. (1.3428, 1.2113).. controls (1.3859, 1.1068) and (1.4462, 1.0176) .. (1.5237, 0.944).. controls (1.601, 0.8705) and (1.6928, 0.8139) .. (1.7992, 0.7744).. controls (1.9053, 0.735) and (2.0214, 0.7152) .. (2.1474, 0.7152).. controls (2.2733, 0.7152) and (2.3892, 0.735) .. (2.4956, 0.7744).. controls (2.6016, 0.8139) and (2.6935, 0.8705) .. (2.771, 0.944).. controls (2.8482, 1.0176) and (2.9086, 1.1068) .. (2.952, 1.2113).. controls (2.9578, 1.2253) and (2.9631, 1.2398) .. (2.9681, 1.2544) -- (2.9681, 0.7562) -- (3.3229, 0.7562) -- (3.3229, 1.4002) -- (3.4981, 1.4002) -- (3.8349, 0.7562) -- (4.2603, 0.7562) -- (3.8505, 1.4364) -- cycle(0.9628, 1.7777).. controls (0.9438, 1.7534) and (0.92, 1.7357) .. (0.8911, 1.7243).. controls (0.8623, 1.7129) and (0.83, 1.706) .. (0.7945, 1.7039).. controls (0.7588, 1.7015) and (0.7252, 1.7005) .. (0.6932, 1.7005) -- (0.4839, 1.7005) -- (0.4839, 2.0667) -- (0.716, 2.0667).. controls (0.7477, 2.0667) and (0.7805, 2.0643) .. (0.8139, 2.0598).. controls (0.8472, 2.0553) and (0.8768, 2.0466) .. (0.9025, 2.0336).. controls (0.9282, 2.0206) and (0.9496, 2.0021) .. (0.9663, 1.9778).. controls (0.9829, 1.9534) and (0.9914, 1.9209) .. (0.9914, 1.8799).. controls (0.9914, 1.8362) and (0.9819, 1.8021) .. (0.9628, 1.7777) -- cycle(2.6125, 1.3533).. controls (2.5889, 1.2904) and (2.5553, 1.2359) .. (2.5112, 1.1896).. controls (2.4672, 1.1433) and (2.4146, 1.1073) .. (2.3529, 1.0814).. controls (2.2916, 1.0554) and (2.2228, 1.0427) .. (2.1471, 1.0427).. controls (2.0712, 1.0427) and (2.0026, 1.0557) .. (1.9412, 1.0814).. controls (1.8799, 1.107) and (1.8272, 1.1433) .. (1.783, 1.1896).. controls (1.7391, 1.2359) and (1.7052, 1.2904) .. (1.6817, 1.3533).. controls (1.6581, 1.4163) and (1.6465, 1.4856) .. (1.6465, 1.5616).. controls (1.6465, 1.6359) and (1.6581, 1.705) .. (1.6817, 1.7687).. controls (1.7052, 1.8325) and (1.7388, 1.8873) .. (1.783, 1.9336).. controls (1.8269, 1.9799) and (1.8796, 2.0159) .. (1.9412, 2.0418).. controls (2.0026, 2.0675) and (2.0712, 2.0804) .. (2.1471, 2.0804).. controls (2.223, 2.0804) and (2.2916, 2.0675) .. (2.3529, 2.0418).. controls (2.4143, 2.0161) and (2.467, 1.9799) .. (2.5112, 1.9336).. controls (2.5551, 1.8873) and (2.5889, 1.8322) .. (2.6125, 1.7687).. controls (2.636, 1.705) and (2.6477, 1.6359) .. (2.6477, 1.5616).. controls (2.6477, 1.4856) and (2.636, 1.4163) .. (2.6125, 1.3533) -- cycle(3.8015, 1.7777).. controls (3.7825, 1.7534) and (3.7587, 1.7357) .. (3.7298, 1.7243).. controls (3.701, 1.7129) and (3.6687, 1.706) .. (3.6333, 1.7039).. controls (3.5975, 1.7015) and (3.5639, 1.7005) .. (3.5319, 1.7005) -- (3.3226, 1.7005) -- (3.3226, 2.0667) -- (3.5547, 2.0667).. controls (3.5864, 2.0667) and (3.6192, 2.0643) .. (3.6526, 2.0598).. controls (3.6859, 2.0553) and (3.7155, 2.0466) .. (3.7412, 2.0336).. controls (3.7669, 2.0206) and (3.7883, 2.0021) .. (3.805, 1.9778).. controls (3.8216, 1.9534) and (3.8301, 1.9209) .. (3.8301, 1.8799).. controls (3.8301, 1.8362) and (3.8206, 1.8021) .. (3.8015, 1.7777) -- cycle;
  \end{scope}
\end{tikzpicture}
}

\usepackage[affil-it]{authblk}
\usepackage{orcidlink}
\author[1%
  *%
  ]{Sherjeel Shabih%
    \,\orcidlink{0009-0008-6635-4465}\,%
    }
\author[1,2%
  *%
  ]{Lukas Pielsticker%
    \,\orcidlink{0000-0001-9361-8333}\,%
    }
\author[1,3%
  ]{Florian Dobener%
    \,\orcidlink{0000-0003-1987-6224}\,%
    }
\author[1%
  ]{Andrea Albino%
    \,\orcidlink{0000-0001-9280-7431}\,%
    }
\author[1%
  ]{Theodore Chang%
    \,\orcidlink{0000-0002-4911-0230}\,%
    }
\author[1,4%
  ]{Carola Emminger%
    \,\orcidlink{0000-0003-4793-1809}\,%
    }
\author[1%
  ]{Lev Ginzburg%
    \,\orcidlink{0000-0001-8929-1040}\,%
    }
\author[1,4%
  ]{Ron Hildebrandt%
    \,\orcidlink{0000-0001-6932-604X}\,%
    }
\author[1%
  ]{Markus Kühbach%
    \,\orcidlink{0000-0002-7117-5196}\,%
    }
\author[1%
  ]{Rubel Mozumder%
    \,\orcidlink{0009-0007-5926-6646}\,%
    }
\author[5%
  ]{Tommaso Pincelli%
    \,\orcidlink{0000-0003-2692-2540}\,%
    }
\author[3%
  ]{Martin Aeschlimann%
    \,\orcidlink{0000-0003-3413-5029}\,%
    }
\author[4%
  ]{Marius Grundmann%
    \,\orcidlink{0000-0001-7554-182X}\,%
    }
\author[2%
  ]{Walid Hetaba%
    \,\orcidlink{0000-0003-4728-0786}\,%
    }
\author[1,6%
  ]{Carlos-Andres Palma%
    \,\orcidlink{0000-0001-5576-8496}\,%
    }
\author[5%
  ]{Laurenz Rettig%
    \,\orcidlink{0000-0002-0725-6696}\,%
    }
\author[1%
  ]{Markus Scheidgen%
    \,\orcidlink{0000-0002-8038-2277}\,%
    }
\author[1%
  ]{José Antonio Márquez Prieto%
    \,\orcidlink{0000-0002-8173-2566}\,%
    }
\author[1%
  ]{Claudia Draxl%
    \,\orcidlink{0000-0003-3523-6657}\,%
    }
\author[1%
  ]{Sandor Brockhauser%
    \,\orcidlink{0000-0002-9700-4803}\,%
    }
\author[1%
  ]{Christoph Koch%
    \,\orcidlink{0000-0002-3984-1523}\,%
    }
\author[7%
  ]{Heiko B. Weber%
    \,\orcidlink{0000-0002-6403-9022}\,%
    }

\affil[1]{Physics Department and CSMB, Humboldt-Universität zu Berlin,
Zum Großen Windkanal 2, D-12489 Berlin, Germany%
    \,\protect\href{https://ror.org/01hcx6992}{\protect\rorlogo}\,%
  }
\affil[2]{Department Heterogeneous Reactions, Max Planck Institute for
Chemical Energy Conversion, Stiftstraße 34-36, D-45470 Mülheim an der
Ruhr, Germany%
    \,\protect\href{https://ror.org/01y9arx16}{\protect\rorlogo}\,%
  }
\affil[3]{Department of Physics, RPTU Kaiserslautern-Landau,
Erwin-Schrödinger-Str. 46, D-67663 Kaiserslautern, Germany%
    \,\protect\href{https://ror.org/01qrts582}{\protect\rorlogo}\,%
  }
\affil[4]{Felix Bloch Institute for Solid State Physics, Leipzig
University, Linnestr. 5, D-04103 Leipzig, Germany%
    \,\protect\href{https://ror.org/03s7gtk40}{\protect\rorlogo}\,%
  }
\affil[5]{Department of Physical Chemistry, Fritz Haber Institute of the
Max Planck Society, Faradayweg 4-6, D-14195 Berlin, DE%
    \,\protect\href{https://ror.org/03k9qs827}{\protect\rorlogo}\,%
  }
\affil[6]{Institute of Physics, Chinese Academy of Sciences, No.8, 3rd
South Street, Zhongguancun, Haidian District, Beijing, China%
    \,\protect\href{https://ror.org/05cvf7v30}{\protect\rorlogo}\,%
  }
\affil[7]{Lehrstuhl für Angewandte Physik,
Friedrich-Alexander-Universität Erlangen-Nürnberg, Staudtstr. 7, D-91058
Erlangen, Germany%
    \,\protect\href{https://ror.org/00f7hpc57}{\protect\rorlogo}\,%
  }
\affil[*]{These authors contributed equally.}
\date{07 July 2025}

\begin{document}
\maketitle

\section{Summary}\label{summary}

Scientific data across physics, materials science, and materials
engineering often lacks adherence to FAIR principles
(\citeproc{ref-Barker:2022}{Barker et al., 2022};
\citeproc{ref-Jacobsen:2020}{Jacobsen et al., 2020};
\citeproc{ref-Wilkinson:2016}{M. D. Wilkinson et al., 2016};
\citeproc{ref-Wilkinson:2025}{S. R. Wilkinson et al., 2025}) due to
incompatible instrument-specific formats and diverse standardization
practices. \texttt{pynxtools} is a Python software development framework
with a command line interface (CLI) that standardizes data conversion
for scientific experiments in materials science to the NeXus format
(\citeproc{ref-Klosowski:1997}{Klosowski et al., 1997};
\citeproc{ref-Koennecke:2006}{Könnecke, 2006};
\citeproc{ref-Koennecke:2015}{Könnecke et al., 2015}) across diverse
scientific domains. NeXus defines data storage specifications for
different experimental techniques through application definitions.
\texttt{pynxtools} provides a fixed, versioned set of NeXus application
definitions that ensures convergence and alignment in data
specifications across, among others, atom probe tomography, electron
microscopy, optical spectroscopy, photoemission spectroscopy, scanning
probe microscopy, and X-ray diffraction. Through its modular plugin
architecture \texttt{pynxtools} provides conversion of data and metadata
from instruments and electronic lab notebooks to these unified
definitions, while performing validation to ensure data correctness and
NeXus compliance. \texttt{pynxtools} can be integrated directly into
Research Data Management Systems (RDMS) to facilitate parsing and
normalization. We detail one example for the RDM system NOMAD. By
simplifying the adoption of NeXus, the framework enables true data
interoperability and FAIR data management across multiple experimental
techniques.

\section{Statement of need}\label{statement-of-need}

Achieving FAIR (Findable, Accessible, Interoperable, and Reproducible)
data principles in experimental physics and materials science requires
consistent implementation of standardized data formats. NeXus provides
comprehensive data specifications for structured storage of scientific
data. \texttt{pynxtools} simplifies the use of NeXus for developers and
researchers by providing guided workflows and automated validation to
ensure complete compliance. Existing solutions
(\citeproc{ref-Jemian:2025}{Jemian et al., 2025};
\citeproc{ref-Koennecke:2024}{Könnecke et al., 2024}) provide individual
capabilities, but none offers a comprehensive end-to-end workflow for
proper NeXus adoption. \texttt{pynxtools} addresses this critical gap by
providing a framework that enforces complete NeXus application
definition compliance through automated validation, detailed error
reporting for missing required data points, and clear implementation
pathways via configuration files and extensible plugins. This approach
transforms NeXus from a complex specification into a practical solution,
enabling researchers to achieve true data interoperability without deep
technical expertise in the underlying standards.

\section{Dataconverter and
validation}\label{dataconverter-and-validation}

The \texttt{dataconverter}, core module of pynxtools, combines
instrument output files and data from electronic lab notebooks into
NeXus-compliant HDF5 files. The converter performs three key operations:
extracting experimental data through specialized readers, validating
against NeXus application definitions to ensure compliance with
existence and format constraints, and writing valid NeXus/HDF5 output
files.

The \texttt{dataconverter} provides a command-line interface (CLI) for
generating NeXus files, supporting both built-in readers for
general-purpose functionality and technique-specific reader plugins,
which are distributed as separate Python packages.

For developers, the \texttt{dataconverter} provides an abstract
\texttt{reader} class for building plugins that process
experiment-specific formats and populate the NeXus specification. It
passes a \texttt{Template}, a subclass of Python's dictionary, to the
\texttt{reader} as a form to fill. The \texttt{Template} ensures
structural compliance with the chosen NeXus application definition and
organizes data by NeXus's required, recommended, and optional levels.

The \texttt{dataconverter} validates \texttt{reader} output against the
selected NeXus application definition, checking for instances of
required concepts, complex dependencies (like inheritance and nested
group rules), and data integrity (type, shape, constraints). It
validates required concepts, reporting errors for any violations, and
issues warnings for invalid data, facilitating reliable and practical
NeXus file generation.

All reader plugins are tested using the \texttt{pynxtools.testing}
suite, which runs automatically via GitHub CI to ensure compatibility
with the dataconverter, the NeXus specification, and integration across
plugins.

The dataconverter includes \texttt{eln\_mapper} that creates either a
fillable \texttt{YAML} file or a \texttt{NOMAD}
(\citeproc{ref-Scheidgen:2023}{Scheidgen et al., 2023}) ELN schema based
on a selected NeXus application definition.

\section{NeXus reader and annotator}\label{nexus-reader-and-annotator}

\texttt{read\_nexus} enables semantic access to NeXus files by linking
data items to NeXus concepts, allowing applications to locate relevant
data without hardcoding file paths. It supports concept-based queries
that return all data items associated with a specific NeXus vocabulary
term. Each data item is annotated by traversing its group path and
resolving its corresponding NeXus concept, included inherited
definitions.

Items not part of the NeXus schema are explicitly marked as such, aiding
in validation and debugging. Targeted documentation of individual data
items is supported through path-specific annotation. The tool also
identifies and summarizes the file's default plottable data based on the
\texttt{NXdata} definition.

\section{\texorpdfstring{\texttt{NOMAD}
integration}{NOMAD integration}}\label{nomad-integration}

While \texttt{pynxtools} works independently, it can also be integrated
directly into any Research Data Management System (RDMS). The package
works as a plugin within the \texttt{NOMAD} platform
(\citeproc{ref-Draxl:2019}{Draxl \& Scheffler, 2019};
\citeproc{ref-Scheidgen:2023}{Scheidgen et al., 2023}) out of the box.
This enables data in the NeXus format to be integrated into
\texttt{NOMAD}'s metadata model, making it searchable and interoperable
with other data from theory and experiment. The plugin consists of
several key components (so called entry points):

\texttt{pynxtools} extends \texttt{NOMAD}'s data schema, known as
\texttt{Metainfo} (\citeproc{ref-Ghiringhelli:2017}{Ghiringhelli et al.,
2017}), by integrating NeXus definitions as a \texttt{NOMAD}
\texttt{Schema\ Package}. This integration introduces NeXus-specific
quantities and enables interoperability by linking to other standardized
data representations within \texttt{NOMAD}. The \texttt{dataconverter}
is integrated into \texttt{NOMAD}, making the conversion of data to
NeXus accessible via the \texttt{NOMAD} GUI. The \texttt{dataconverter}
also processes manually entered \texttt{NOMAD} ELN data in the
conversion.

The \texttt{NOMAD} Parser module in \texttt{pynxtools}
(\texttt{NexusParser}) extracts structured data from NeXus HDF5 files to
populate \texttt{NOMAD} with \texttt{Metainfo} object instances as
defined by the \texttt{pynxtools} schema package. This enables ingestion
of NeXus data directly into \texttt{NOMAD}. Parsed data is
post-processed using \texttt{NOMAD}'s \texttt{Normalization} pipeline.
This includes automatic handling of units, linking references (including
sample and instrument identifiers defined elsewhere in \texttt{NOMAD}),
and populating derived quantities needed for advanced search and
visualization.

\texttt{pynxtools} contains an integrated \texttt{Search\ Application}
for NeXus data within \texttt{NOMAD}, powered by \texttt{Elasticsearch}
(\citeproc{ref-elasticsearch:2025}{Elasticsearch B.V., 2025}). This
provides a search dashboard whereby users can efficiently filter
uploaded data based on parameters like experiment type, upload
timestamp, and domain- and technique-specific quantities. The entire
\texttt{pynxtools} workflow (conversion, parsing, and normalization) is
exemplified in a representative \texttt{NOMAD} \texttt{Example\ Upload}
that is shipped with the package. This example helps new users
understand the workflow and serves as a template to adapt the plugin to
new NeXus applications.

\section{Funding}\label{funding}

The work is funded by the Deutsche Forschungsgemeinschaft (DFG, German
Research Foundation) - project 460197019 (FAIRmat).

\section{Acknowledgements}\label{acknowledgements}

We acknowledge the following software packages our package depends on:
\texttt{h5py} (\citeproc{ref-H5py:2008}{Collette \& al., 2008}),
\texttt{numpy} (\citeproc{ref-Harris:2020}{Harris et al., 2020}),
\texttt{click} (\citeproc{ref-Click:2014}{The Pallets development team,
2014}) , \texttt{CFF} (\citeproc{ref-Druskat:2021}{Druskat et al.,
2021}), \texttt{xarray} (\citeproc{ref-Hoyer:2017}{S. Hoyer \& Hamman,
2017}), (\citeproc{ref-Hoyer:2025}{Stephan Hoyer et al., 2025}),
\texttt{pandas} (\citeproc{ref-Pandas:2020}{The pandas development team,
2020}), \texttt{scipy} (\citeproc{ref-McKinney:2010}{McKinney, 2010}),
\texttt{lxml} (\citeproc{ref-Behnel:2005}{Behnel et al., 2005}),
\texttt{mergedeep} (\citeproc{ref-Clarke:2019}{Clarke, 2019}),
\texttt{Atomic\ Simulation\ Environment}
(\citeproc{ref-Hjorth:2017}{Larsen et al., 2017}), \texttt{pint}
(\citeproc{ref-Pint:2012}{The Pint development team, 2012}).

\section*{References}\label{references}
\addcontentsline{toc}{section}{References}

\phantomsection\label{refs}

\end{document}